\documentclass[aps,prb,floatfix,twocolumn,showpacs,preprintnumbers,amsmath,amssymb,superscriptaddress]{revtex4}

\usepackage{graphicx}  % Include figure files
\usepackage{dcolumn}   % Align table columns on decimal point
\usepackage{bm}        % bold math
\usepackage{stmaryrd}  % for mapsfrom symbol
\usepackage{multirow}  % multirow tabular
\usepackage{color}
\newcommand{\der}[3]{\frac{{\rm d}^{#1} #2}{{\rm d} #3^{#1}}}     % n-th derivation
\newcommand{\av}[1]{ {\langle #1 \rangle} }                       % average value

\def   \bhs      {{\hat{\bm s}}}                         % unit-length sensing-layer spin
\def   \bhS      {{\hat{\bm S}}}                         % unit-length reference-layer spin
\def   \Ms       {M_{\rm s}}                             % saturated magnetization
\def   \Heff     {{\bm H}_{\rm eff}}                     % effective field
\def   \hext     {H_{\rm ext}}                           % external field value
\def   \hani     {H_{\rm ani}}                           % anisotropy field value
\def   \Hdemag   {{\bm H}_{\rm dem}}                     % demagnetization field
\def   \Hth      {{\bm H}_{\rm th}}                      % thermal mag. field
\def   \ex       {{\hat{\bm e}_{x}}}                     % unit vector along the x-axis
\def   \ez       {{\hat{\bm e}_{z}}}                     % unit vector along the z-axis
\def   \tor      {\bm\tau}                               % spin-torque
\def   \tin      {\bm\tau_\theta}                        % in-plane component of spin-torque
\def   \tout     {\bm\tau_\varphi}                       % out-of-plane component of spin-torque
\def   \ns       {{\rm ns}}                              % nanosecond
\def   \Oe       {{\rm Oe}}                              % oersted
\def   \K        {{\rm K}}                               % Kelvin
\def   \ts       {t_{\rm s}}                             % Switching time
\def   \tp       {t_{\rm p}}                             % Pulse duration
\def   \ema      {{\overline{s_{z}}}}                    % Exponentially weighted moving average
\def   \sx       {{s_{x}}}
\def   \sy       {{s_{y}}}
\def   \sz       {{s_{z}}}
\def   \SSp      {{\rm SS}_{+}}                          % stable state close to +ex
\def   \SSm      {{\rm SS}_{-}}                          % stable state close to -ex
\def   \SSpm     {{\rm SS}_{\pm}}                        % both stable states
\def   \Psw      {P_{\rm sw}}                            % Switching probability

\begin{document}

\preprint{APS/123-QED}

\title{Current-pulse induced magnetic switching in standard and nonstandard spin-valves}

\author{P.~Bal\'a\v{z}}
\email{balaz@amu.edu.pl}
\affiliation{Department of Physics, Adam Mickiewicz University,
             Umultowska 85, 61-614~Pozna\'n, Poland}
\author{M.~Gmitra}
%\email{martin.gmitra@upjs.sk}
\affiliation{Institute of Phys.,
             P. J. \v{S}af\'arik University, Park Angelinum 9,
             040 01 Ko\v{s}ice, Slovak Republic}
\author{J.~Barna\'s}
%\email{barnas@amu.edu.pl}
\affiliation{Department of Physics, Adam Mickiewicz University,
             Umultowska 85, 61-614~Pozna\'n, Poland}
\affiliation{Institute of Molecular Physics, Polish Academy of Sciences
              Smoluchowskiego 17, 60-179 Pozna\'n, Poland}

\date{\today}

\begin{abstract}
Magnetization switching due to a current-pulse in symmetric and
asymmetric spin valves is studied theoretically within the
macrospin model. The switching process and the corresponding
switching parameters are shown to depend significantly on the
pulse duration and also on the interplay of the torques due to
spin transfer and external magnetic field. This interplay leads to
peculiar features in the corresponding phase diagram. These
features in standard spin valves, where the spin transfer torque
stabilizes one of the magnetic configurations (either parallel or
antiparallel) and destabilizes the opposite one, differ from those
in nonstandard (asymmetric) spin valves, where both collinear
configurations are stable for one current orientation and unstable
for the opposite one. Following this we propose a scheme of
ultrafast current-induced switching in nonstandard spin valves,
based on a sequence of two current pulses.
\end{abstract}

\maketitle

%----------------------------------------------
\section{Introduction}
%----------------------------------------------

% Introduction to CIMS
The possibility of current-induced magnetic switching (CIMS) in
multilayer structures has been introduced by Slonczewski
\cite{Slonczewski1996:JMMM} and Berger \cite{Berger1996:PRB}. They
pointed out that spin polarized current passing through a
multilayer consisting of two ferromagnetic layers separated by a
nonmagnetic spacer can exert a torque on the local magnetic
moments in ferromagnetic layers. In case of commonly used
(standard) spin valves, this torque can switch the system between
parallel (P) and antiparallel (AP) alignments of the layers'
magnetic moments. Indeed, the CIMS has been confirmed
experimentally in many spin valve structures
\cite{Tsoi1998:PRL,Myers1999:Science,Katine1999:PRL,
Kiselev2003:Nature}.

In order to meet the technological demands for applications, 
designing of  bistable magnetic spin valve devices
that could be fast switched by electric current only is highly
desired. Accordingly, several subtle switching schemes aimed at
speeding up the switching process and lowering the energy costs have
been developed for various magnetic multilayer structures. It has
been shown that proper optimization of the current pulse
parameters (amplitude, duration, frequency, etc.) can finally
result in ultrafast, single-step, switching
\cite{Serrano-Guisan2008:PRL,Garzon2008:PRB}.

Current-induced dynamics in spin valves is governed by
spin-transfer torque (STT), particularly by its magnitude and
dependence on the angle between magnetization vectors of the
reference and sensing magnetic layers. Such an angular variation
depends mainly on the type of electron transport through the spin
valve. Recently, STT in the diffusive transport regime draws more
attention. It has been demonstrated \cite{Urazhdin2008:PRB}, that
spin diffusion in various types of polarizing layer can strongly
influence the current-induced behavior of spin valves. To
calculate  STT in the diffusive transport limit, one can consider
the model \cite{Barnas2005:PRB} that extends the Valet-Fert
description \cite{Valet1993:PRB} to arbitrary magnetic
configuration and allows to study STT as a function of layers'
thicknesses, type of materials, spin asymmetries, etc, which can
be directly compared to experimental observations
\cite{Urazhdin2003:PRL}. In the case of diffusive transport,
spin-diffusion length has to be much longer than the mean free
path. However, it has been shown \cite{Penn2005:PRB} that
diffusive approach is valid also for spin diffusion lengths of the
order of mean free paths.  Moreover, the approach describes
relatively well experimental results even for layer thicknesses
comparable to the corresponding mean free paths.

One of the most pronounced manifestation of the importance of
spin-diffusion length has been shown for asymmetric spin valves
\cite{Barnas2005:PRB}, where the non-standard (wavy-like) angular
dependence of  STT leads to the current-induced precessional
regimes in zero magnetic field \cite{Gmitra2006:PRL,
Gmitra2006:APL,Gmitra2007:PRL}. This prediction has been later
confirmed experimentally \cite{Boulle2007:NP,Boulle2008:PRB}. The
wavy-like STT vanishes not only in the collinear configurations --
P and AP ones -- but also in a certain non-collinear
configuration. Depending on the current direction,  the collinear
configurations are either both stable or both unstable. The first
case is of importance for the current-induced zero-field microwave
excitations, while the later one would be desired for further
stabilization of the collinear magnetic configurations in memory
devices against thermal or current fluctuations. Since the problem
of switching in non-standard spin valves has not been addressed
yet, here we present a comprehensive study of the CIMS in such
systems due to a current pulse of finite duration. In the
following, the spin valves with wavy-like STT will be refereed to
interchangeably as nonstandard or asymmetric ones, whereas the
spin valves with STT vanishing only in collinear configurations
will be referred to as standard or symmetric ones.

In this paper, we investigate current-pulse-induced magnetization
dynamics in three-layer spin valves ${\rm X/Cu(10)/Py(8)}$
sandwiched between semi-infinite Cu leads. The numbers in brackets
correspond to the layers' thicknesses in nanometers. The Py(8)
stands for the permalloy sensing layer, which magnetization
direction can freely rotate upon applied magnetic field and/or
spin polarized current, whereas ${\rm X}$ denotes the reference
layer, which magnetization is considered to be fixed and not
influenced by external magnetic fields and electric current.
Assuming these conditions, we have considered two different types
of the reference layer: ${\rm X=Py(20)}$, and ${\rm X=Co(8)}$.
Taking into account the above introduced notation,  the former
spin valves, ${\rm Py/Cu/Py}$, we will referred to as standard or
symmetric, whereas the latter spin valves, ${\rm Co/Cu/Py}$, we
will be referred to as non-standard or asymmetric ones.

The main motivation of this paper is to provide a systematic study
of the CIMS, based on the macrospin simulations. We consider a
rectangular current pulse characterized by the current amplitude
and pulse duration. We have found that a proper choice of the
pulse parameters leads to fast switching with relatively low
energy costs. In the case of non-standard spin valves, the
double-pulse scheme, which allows reliable switching
between the collinear magnetic configurations and significant
shortening of the overall switching time, is proposed.

The paper is organized as follows. In section~2 we describe the
macrospin model. Numerical analysis for both standard and
non-standard (asymmetric) spin valve is described in section~3,
where also a double-pulse switching scheme for an asymmetric spin
valve is described. Section~4 includes summary and final
conclusions.

%----------------------------------------------
\section{Macrospin model}
%----------------------------------------------

Time resolved imaging of CIMS showed that inhomogeneous
spatio-temporal magnetization evolution takes place during the
reversal \cite{Acremann2006:PRL}. Thus, more sophisticated models,
involving detailed micromagnetic description, have to be adopted
for a better qualitative description \cite{Berkov2005:PRB}.
However, the macrospin model -- particularly for spin valves in
nano-size range --  provides a sufficient quantitative description
of the current-induced dynamics, which is in agreement with many
experiments on standard spin valves \cite{Kiselev2003:Nature}. A
relatively good qualitative agreement has been reached also for
non-standard (asymmetric) spin valves, where some predictions
based on the macrospin model \cite{Gmitra2006:PRL,
Gmitra2006:APL,Gmitra2007:PRL} have been confirmed experimentally
\cite{Boulle2007:NP,Boulle2008:PRB}.

Time evolution of the unit vector $\bhs=(\sx, \sy, \sz)$ along the
net spin moment of the sensing layer is described by the
generalized Landau-Lifshitz-Gilbert equation,
\begin{equation}
\label{Eq:LLG}
  \der{}{\bhs}{t} = - |\gamma_{\rm g}| \mu_0 \, \bhs \times \Heff
                    - \alpha \, \bhs \times \der{}{\bhs}{t}
                    + \frac{|\gamma_{\rm g}|}{\Ms d} \, \tor \, ,
\end{equation}
where $\gamma_{\rm g}$ is the gyromagnetic ratio, $\mu_0$ is the
magnetic vacuum permeability, $\Ms$ is the saturation
magnetization and $d$ is the sensing layer thickness. The Gilbert
damping parameter $\alpha$ is assumed to be constant, $\alpha =
0.01$. The effective field $\Heff$ includes contributions from the
external magnetic field ($\hext$), uniaxial magnetic anisotropy
($\hani$), demagnetization field ($\Hdemag$) and the thermal field
($\Hth$); $\Heff = -\hext \ez - \hani \left(\bhs \cdot \ez \right)
\ez + \Hdemag + \Hth$, where $\ez$ is the unit vector along the
axis $z$ which is parallel to the in-plane magnetic easy axis. By
definition, the external field is positive when it is pointing
along the negative $z$ axis. The demagnetization field corresponds
to the sensing layer of an elliptical shape with the major and
minor axis of 130 and 60 nanometers, respectively, and thickness
of 8 nanometers. The magnetic easy axis is assumed to be along the
longer axis of the ellipse, and $\hani=100.5\,{\rm Oe}$.

The thermal field $\Hth$ contributing to $\Heff$ is a stochastic
field of statistical properties: $\av{H_{{\rm th},\, i}(t)}=0$ and
$\av{ H_{{\rm th},\, i}(t) H_{{\rm th},\, j}(t') } = 2 {\cal D}
\delta_{ij} \delta(t-t')$, where $i,\, j \in \{x,y,z\}$
The strength of the thermal fluctuations is given by the parameter
${\cal D} = 2 k_{\rm B} T / [\Ms (1 + \alpha^{2})]$ derived
according to the fluctuation-dissipation relation
\cite{Brown1963:PR,Garcia-Palacios1998:PRB}, with $T$ being the
temperature.

As concerns the STT, we consider both the in-plane and
out-of-plane components, $\tor = \tin + \tout$, where $\tin = a I
\, \bhs \times (\bhs \times \bhS)$ and $\tout = b I \, \bhs \times
\bhS$. The vector $\bhS$ points along the net spin of the
reference layer ${\rm X}$, $\bhS =\ez$, and is assumed to be
constant in time (current does not excite magnetic moment of the
reference layer). Current density $I$ is defined as positive when
current flows from the reference layer towards the sensing one.
Finally, the angular dependence of the parameters $a$ and $b$ has
been calculated in the diffusive transport limit
\cite{Barnas2005:PRB}. The resulting in-plane and out-of-plane
torques for the spin valves ${\rm X/Cu(10)/Py(8)}$ with ${\rm
X=Py(20)}$ and ${\rm Co(8)}$ are shown in Fig.~1 as a function of
the angle $\theta$ (azimuthal angle) between the magnetic moments
of the reference and sensing layers ($\bhs\cdot\bhS=\cos\theta$).
%------------------------------------------------------------------------------
\begin{center}
  \begin{figure}[t]
    \label{torques}
    \includegraphics[angle=0,width=0.99\columnwidth]{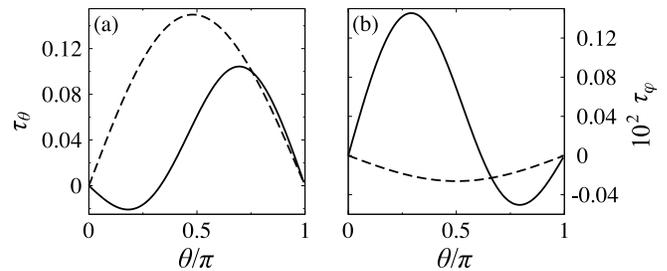}
    \caption{
Angular dependence of the spin transfer torque components in units of $\hbar I / |e|$ 
acting on th Py(8) in the ${\rm Cu/X/Cu(10)/Py(8)/Cu}$ spin valves for
the standard case with ${\rm X=Py(20)}$ (dashed lines) and
nonstandard case (wavy-like torque)  with ${\rm X=Co(8)}$ (solid
lines). (a)~In-plane component $\tau_\theta$, and (b)~out-of-plane
component $\tau_\varphi$. The other parameters as in
Refs~[\onlinecite{Barnas2005:PRB,Gmitra2007:PRL}].}
  \end{figure}
\end{center}
%------------------------------------------------------------------------------

%----------------------------------------------
\section{Numerical results}
%----------------------------------------------
Here we study switching in the ${\rm X/Cu(10)/Py(8)}$ valves from
the P to AP state due to a current pulse. Numerical solution of the %Eq.(\ref{Eq:LLG})
LLG equation has been performed using the Heun Scheme
\cite{Kloeden1992:Springer} with an auto-adaptive time step. For
the initial configuration we assume a biased state with $\theta
(t=0) = 1^{\circ}$ and $\varphi (t=0) = \pi / 2$, where $\varphi$
is the polar angle describing orientation of the vector $\bhs$
from the $x$ axis parallel to the current flow. The current pulse
$i(t)$ of constant current density $I$ and duration $\tp$ is
applied at $t=0$, $i(t)=I[\Theta (t)-\Theta(t-t_p)]$, where
$\Theta (x)=1$ for $x>0$ and $\Theta (x)=0$ for $x\le 0$. A
successful switching event with the corresponding switching time
$\ts$ is counted when $\ema (\ts) < -0.99$, where $\ema(t)$ is the
exponentially weighted moving average \cite{Roberts2000}, $\ema(t)
= \eta~ \sz (t) + (1 - \eta) \ema (t - \Delta t)$, $\Delta t$ is
the integration step, and the weighting parameter $\eta = 0.1$.
The moving average $\ema$ is calculated for time $t>t'$ when
$\sz(t')$ reaches the value of $-0.9$; otherwise $\ema(t) =
s_z(t)$. From the experimental point of view it is more convenient
to reformulate the switching condition in terms of the
magnetoresistance. We note that this holds only for the standard
spin valves, in which the magnetoresistance is a monotonic
function of the $\theta$ angle \cite{Bauer2003:PRB,
Gmitra2009:PRB}. In asymmetric spin valves the magnetoresistance
can be a non-monotonic function of $\theta$ \cite{Gmitra2009:PRB,
Urazhdin2005:PRB}, and therefore direct calculation of the
magnetoresistance is then needed.

%----------------------------------------------
\subsection{Dynamics in a Py/Cu/Py spin valve}
%----------------------------------------------

Let us study first the standard spin valve, X=Py(20). The in-plane
and out-of-plane components of the STT acting on the Py(8) sensing
layer show sine-like angular dependences, see dashed lines in
Fig.~1. Switching time as a function of the pulse duration $\tp$
and reduced current density $I/I_0$ ($I_0 = 10^8 {\rm A cm}^{-2}$)
is shown in Fig.~2(a). Here, we consider zero temperature limit and
the external magnetic field is set to zero. Two different regions
in the switching diagram can be distinguished. First, the white
non-switching region is observed for short current pulses and low
current densities. In this region, the energy gain due to STT does
not overcome the Gilbert damping and system stays in the
initial local magnetic energy minimum. The second region
corresponds to successful switching to the AP state. The switching
time $\ts$ is shown in the color scale. The $\ts$ decreases
nonmonotonously with increasing current density. The most bright
area corresponds to the ultra-fast switching, in which the spin
reaches the AP configuration before the current pulse ends ($\ts <
\tp$). In such a case the switching is realized in a single
ultra-fast step after a half precession around the $x$ axis; see
Fig.2(b) right.

The boundary between non-switching and ultra-fast switching
regions develops into a ripple structure. In this region, the
energy gain due to spin-transfer leads to a {\em retarded
switching}, where the switching time $\ts > \tp$, see Fig.2(b)
left. The switching for $t > \tp$ is accompanied with a ringing,
where the spin relaxes to the AP state due to energy dissipation
via the Gilbert damping only. Such dissipation, however, is rather
slow and therefore the retarded switching is much more slower than
the ultra-fast single-step switching.

%------------------------------------------------------------------------------
\begin{center}
  \begin{figure}[h!]
    \label{PyCuPy_sd}
    \includegraphics[angle=0,width=0.99\columnwidth]{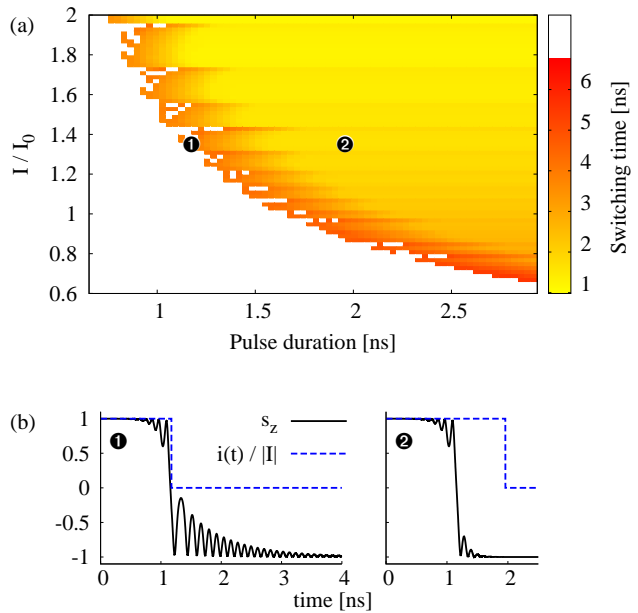}
    \caption{(Color online)
      Current-pulse driven switching in the Py(20)/Cu(10)/Py(8) spin
      valve in the absence of magnetic field. (a)~Switching time as a
      function of the pulse duration $\tp$ and reduced current density $I/I_0$,
      $I_0 = 10^8 {\rm A cm}^{-2}$.
      (b)~Temporal evolution of the $s_z$ spin
      component under rectangular current pulse (dashed lines) of
      amplitude $I/I_0 = 1.4$ and duration $\tp=1.2\,{\rm ns}$ ({\em
      retarded switching}, left) and $\tp=2\,{\rm ns}$ ({\em fast
      switching}, right),  corresponding to the points marked in the switching diagram (a).}
  \end{figure}
\end{center}
%------------------------------------------------------------------------------

To speed up switching from P to AP state one may consider a
negative external magnetic field. On the other hand, a positive
magnetic field exceeding the anisotropy field leads to commonly
observed steady-state out-of-plane precessional (OPP)
modes~\cite{Myers1999:Science,Katine1999:PRL,Kiselev2003:Nature},
which are the result of the energy balance between Gilbert damping
and the energy gain due to spin-transfer.

Our analysis shows that for positive external magnetic field, the
continuous switching region in the diagram shown in Fig.~2(a)
splits into an non-compact current dependent stripe structure. In
Fig.~3(a) we show the switching diagram for $\hext=200\,{\rm Oe}$.
The switching regions alternate with the stripes where the spin
transfer induces the OPP regime. Since, the current pulse is
finite, the final state depends on the actual spin state at the
$t=\tp$, which falls into the basin of attraction either of P or
AP state. This is shown in Fig.~3(b), where two switching events
under the current pulses of the same amplitude ($I = 2.75\, I_0$)
and different pulse duration are driven via the OPP regime. We
note that in the switching regions the spin dynamics is similar to
the zero-field switching discussed above.

To elucidate the stripe structure, we plotted in Fig.~3(c) map
of the final spin states as a function of the initial spin position
$\bhs_0 = \bhs (t=0)$, assuming constant current amplitude and
pulse duration $\tp \to \infty$. The gray (black) regions
correspond to the initial spin position, which results in the
final OPP (AP) regime. Comparing the maps calculated for two
different current densities ($I = 2.75\, I_0$ and $I =  3.00\,
I_0$), one concludes that the dynamical phase portrait depends
rather strongly on the current density. In other words,
current-driven dynamics from the same initial state can develop to
different final states. Further increase of external magnetic
field leads to shrinking of the P$\to$AP switching stripes. For
fields much larger than the coercive field the switching stripes
disappear so the OPP regime remains only. We note that the initial
spin position (close to the P configuration) assumed in this paper
(except Fig.3(c)) is denoted in Fig.3(c) by the circles.

The sharp stripe structure is a result of deterministic dynamics
and fixed initial condition. When a distribution of initial
configurations is taken into account, some  smearing of the border
between the stripes is observed (not shown). The boundaries are
also smeared  when non-zero temperature is considered. In
Fig.~3(d) we show  the switching probabilities as a function of
the current pulse density, calculated for pulse duration $\tp =
3\,\ns$ at  $T = 4.2\,\K$ and $T = 77\,\K$, and  for fixed initial
configuration. The statistics has been calculated from $10^4$
events for each value of the current density. The switching
probability follows the stripe structure in the zero temperature
limit, and decreases with decreasing current amplitude. For $T =
77\,\K$, the probability is lowered by the factor of about 3 and
the peaks broaden. From this follows that smearing of the
boundaries between switching and precessional regions increases
with increasing temperature. In addition, positions of the peaks
are shifted, which reflects the fact that thermal fluctuations act
on the spin like an additional torque, and non-linearly influence
the spin dynamics.
%------------------------------------------------------------------------------
\begin{center}
  \begin{figure}[h]
    \label{PyCuPy_sdh}
    \includegraphics[angle=0,width=0.99\columnwidth]{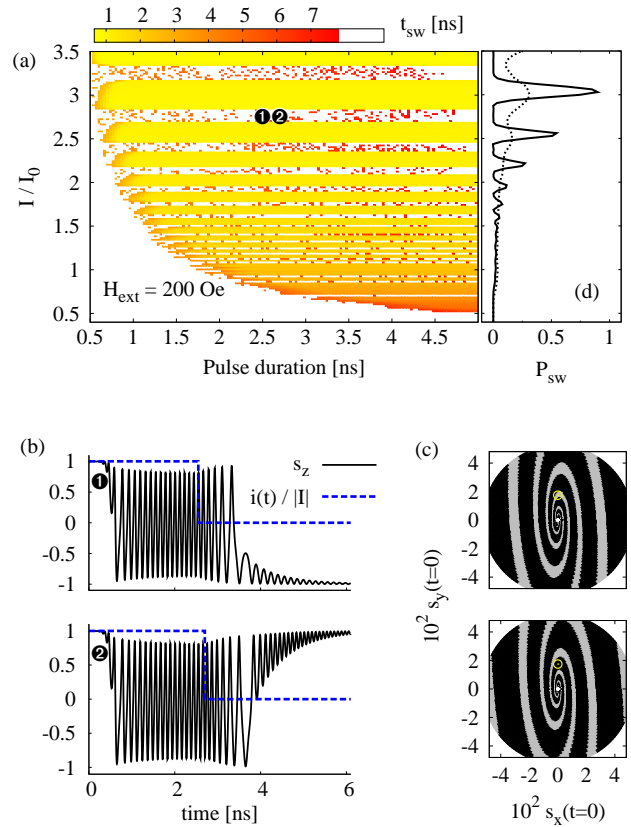}
    \caption{(Color online)
      Effect of the external magnetic field $\hext=200\,\Oe$ on the
      current-pulse driven dynamics in the Py(20)/Cu(10)/Py(8) spin
      valve.
      (a)~Switching time as a function of
      reduced current density $I/I_0$ ($I_0 = 10^8 {\rm A cm}^{-2}$)
      and pulse duration $\tp$.
      (b)~Temporal evolution of
      the $\sz$ spin component under current pulses (dashed lines)
      marked in the switching diagram (a), corresponding to the
      amplitude $I=2.75\, I_0$ and durations
      $\tp = 2.56\,{\rm ns}$ (upper part) and
      $\tp = 2.7\,{\rm ns}$ (lower part).
      (c)~The maps of the final states as a function of initial spin bias
      for current pulse of $I = 2.75\, I_0$ (upper part) and
      $I = 3.0\, I_0$ (lower part), and for $\tp \to \infty$.
      The spin dynamics initialized from the points inside the
      gray (black) areas leads finally to OPP regime (AP state).
      Circles denote the initial spin bias used in the simulations.
      (d)~Thermally assisted switching probability, $\Psw$,  from P to AP state,
      driven by a $3\,\ns$ rectangular current pulse,
      calculated as a function of current density at
      $T= 4.2\,\K$ (solid line) and
      $T= 77\,\K$ (dotted line).}
  \end{figure}
\end{center}
%------------------------------------------------------------------------------

%----------------------------------------------
\subsection{Dynamics in a Co/Cu/Py spin valve}
%----------------------------------------------

The Co(8)/Cu(10)/Py(8) spin valve exhibits non-standard STT acting
on the Py(8) layer. Due to the {\it wavy}-like dependence of the
STT, shown by the solid lines in Fig.~1, positive current
stabilizes both the P and AP configurations. A negative current,
in turn, destabilizes both the collinear configurations. This
characteristic property of the {\em wavy} torque raises the
question, whether it is possible to switch an asymmetric spin
valve between P and AP states without the need of an external
magnetic field.

In Fig.~4(a) we show the switching diagram from P to AP state
under a rectangular current pulse. Here one may distinguish four
characteristic switching regions. First region, denoted by (i),
corresponds to low current amplitudes and/or short pulses, where
switching does not take place. The non-compact region (ii) of
rib-like structure borders the non-switching region and comprises
relatively short pulses leading to the fast switching processes.
In region (iii) the P/AP bistability of the final states is
observed. Finally, in the region (iv) the final state of the
dynamics depends on the applied current density, resulting in the
band-like structure. The structure contains regions with final P
state, which regularly alternate with the regions of final AP
state.

In order to explain the complex diagram structure, let us study
current-pulse-induced dynamics due to the $6~{\rm ns}$ pulse of
amplitude $I=-3.45\, I_0$ at zero temperature. The temporal
dependence of the spin components is shown in Fig.~4(b). When the
constant current pulse is applied (in zero external magnetic
field), it induces initially small-angle in-plane precessions
(IPP) of the sensing layer around the $z$-axis. The precessional
angle rapidly increases and spin dynamics turns to the OPP regime,
where spin precesses mainly around the demagnetization field. The
out-of-plane component of the STT, $\tout$, assists in the
transition to the OPP-like regime \cite{Gmitra2006:PRL}. Numerical
analysis reveals that the transition depends on the current
amplitude, and spin can precess with positive or negative $\sx$
component. Considering constant initial spin direction, OPP
direction depends mainly on the current density. This appears
because the spin phase portrait, and hence the spin trajectory,
are modified due to the current density. Such a situation is
similar to that discussed in Fig.~3(c) for the Py/Cu/Py spin
valve. Due to the sustained energy pumping to the system via the
spin-transfer, the OPP angle decreases and the spin is finally
driven into one of the possible static states (SS) close to the $\ex$
($\SSp$), or $-\ex$ ($\SSm$), depending on the sign of the $\sx$
component in the OPP regime. The $\SSpm$ states are the static
fixed points that result from the interplay between the STT and
effective magnetic field (mainly its demagnetization part). The
$\SSpm$ points are close to the maximum magnetic energy.
Therefore, if current is turned off, the spin position becomes
unstable; spin is driven due to Gilbert damping through the OPP
regime with decreasing precessional frequency to the IPP regime.
In the IPP regime, spin precesses around $+\ez$ ($-\ez$) direction
and is finally damped to the P (AP) state. We have observed that
position of the spin in the $\SSm$ ($\SSp$) results in the final P
(AP) state, see Fig.~4(b). Thus the alternation between P and AP
states in the region (iv) in the diagram shown in Fig.~4(a), is
predominantly controlled by the current since the position of the
static state depends on the current density.
%------------------------------------------------------------------------------
\begin{center}
  \begin{figure}[!th]
    \label{CoCuPy_sd}
    \includegraphics[angle=0,width=0.99\columnwidth]{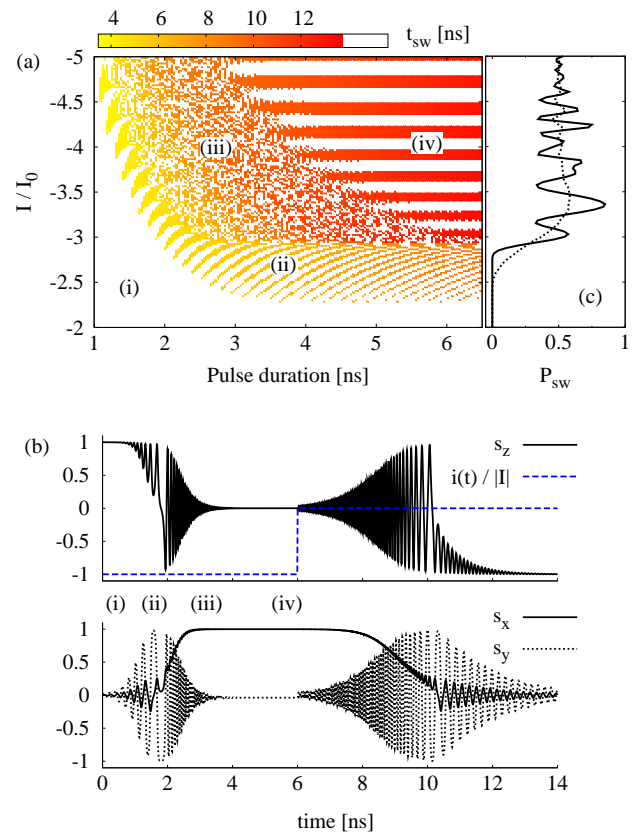}
    \caption{(Color online)
      Current-pulse induced switching in the asymmetric Co(8)/Cu(10)/Py(8)
      spin valve at zero magnetic field.
      (a)~Switching time as a function of the pulse duration $\tp$ and
      reduced current density $I/I_0$,
      $I_0 = 10^8 {\rm A cm}^{-2}$.
      (b)~Temporal evolution of $\bhs$ under the $6\,\ns$ current pulse
      of the amplitude $I = -3.45\, I_0$.
      (c)~Thermally assisted switching probability, $\Psw$, from P to AP,
      driven by $6\,\ns$ rectangular current pulse,
      calculated as a function of current density at
      $T = 4.2\,\K$ (solid line) and $T = 77\,\K$ (dotted line).}
  \end{figure}
\end{center}
%------------------------------------------------------------------------------

To elucidate other regions in the diagram [Fig.~4(a)], we have to
consider shorter pulses. According to the diagram, to have a
successful switching event, the pulse has to exceed a critical
current density and duration. In the static limit ($\tp\to\infty$)
the critical density is about $I = -1.0\, I_0$. For a finite pulse
duration, higher densities are necessary to drive the spin during
the time $\tp$ away from the P state. When the pulse is shorter
than a critical one, the relaxation back to the P state takes
place [region (i)]. To escape the basin of attraction, the time
$\tp$ has to exceed an escape time that depends on the actual
magnetic energy and the current density. For the higher current 
densities, the spin is driven faster away and a shorter escape time is
needed. If the pulse ends just before the onset of the OPP regime,
the spin is then placed within the basin of attraction of the AP
state. The relaxation via the Gilbert dissipation drives the
system to the AP state [region (ii)]. If the spin is driven
further away from the P state, the IPP regime switches fast to the
OPP one. In such a case the final state strongly depends on the
precession phase at $t=\tp$, i.e., when the current is turned off.
This gives rise to the P/AP bistability in the region (iii). The
bistable regime appears up to the pulse duration that is not
longer than the time necessary for spin stabilization in the
$\SSpm$ state. Note, that this analysis is valid only for current
amplitudes $I \gtrsim 3\, I_0$. In case of smaller amplitudes,
only the steady-state large-angle IPP regime has been observed,
and apart from the region (i)  only the region (ii) is present.
The periodic rib-like structure in the region (ii) arises from the
dependence of the final state on the precession phase at time
$t=\tp$.

When temperature is nonzero, the final state is affected by the
thermal noise that modifies the overall spin switching trajectory.
In Fig.~4(c) we show switching probability as a function of the
current density for $\tp = 6\,\ns$, and $T = 4.2\,\K$ (solid line)
and $T = 77\,\K$ (dotted line). The switching probability
oscillates following the zero-temperature stripe structure,
similarly as for the spin valve Py(20)/Cu(10)/Py(8), see
Fig.~3(b). However, the probability $\Psw$ oscillates now around
the value of $\Psw =0.5$, and for increased current densities
approaches this value for any temperature. We have found that for
higher current densities the spin is driven via the OPP-like
transient regime much closer to the SS state. This regime is
sensitive to the thermal fluctuations mainly due to the component
of the thermal field transverse to the spin trajectory. When the
spin remains in the transient regime for a longer time, the impact
of the thermal fluctuations is larger and leads to equilibration of
the probabilities for switching to the P and AP states ($\Psw
\rightarrow 0.5$).

%------------------------------------------------
\subsection{Switching in nonstandard spin valves}
%------------------------------------------------

The fastest switching process in the asymmetric spin valves
appears in the region (ii) [see Fig.4(a)]. This region, however,
is non-compact and therefore to obtain a successful switching one
has to set the current pulse parameters very precisely. In the
region (iii) the bistability of the final state makes the
switching out of control. Thus, the most convenient for switching
seems to be the region (iv). For a proper choice of parameters
(pulse duration and current amplitude, including also the thermal
effects), corresponding to the maximum of the $\Psw$, see Fig.~4(c),
it is possible to obtain controllable switching. However, complex
spin dynamics, especially the ringing which appears after the end
of current pulse, significantly lengthen the switching time. In
practice, the longer the switching time is the more sensitive is
the spin evolution to the external disturbances  and temperature.
Therefore, it is highly desired from the applications point of
view to shorten the switching time as much as possible.
Accordingly, we propose here a double-pulse switching scheme. The
scheme includes two rectangular current pulses of certain
amplitudes and durations. The first pulse of negative current,
referred to as {\it destabilizing pulse}, drives the spin out of
its initial position. We note that  both the collinear
configurations are now unstable. The second pulse of positive
current, called {\it stabilizing pulse},  controls the dynamics
and drives the spin into the final state. Moreover, the
stabilizing pulse shortens the switching time (suppressing the
ringing) via additional energy dissipation from the system.

In Fig.~5(a) we show time evolution of the $\sz$ component due to
single current-pulse and double current-pulse. Here we consider
infinitely long stabilizing pulse that in principle has no effect
on the spin dynamics for $t \gtrsim 6.5\,\ns$. Due to the first
current pulse of density $I = -4.0\,I_0$, the STT drives the spin
to the SS state. When the first pulse is not followed by the
second (stabilizing) one, the spin returns back via the OPP and
IPP regimes to the initial state.
In the case of double-pulse, however, the stabilizing pulse of $I
= 2.0\,I_0$ drives the spin to the AP state. More systematical
study reveals that including the stabilizing current pulse of $I =
2\, I_0$ leads to considerable modification of the switching
diagram (not shown). More specifically, the region (ii) becomes
wider and switching times under this current pulses falls down
from $4$  to $1\; \ns$. Bistability in the region (iii) becomes
reduced, but still not completely removed. Finally, in the region
(iv) the bands related to switching become enlarged, e.g., at $T =
0\,\K$, in the range of amplitudes from $I \simeq -3.5\, I_0$ to $
-4.5\, I_0$ one obtains controllable switching for pulses $\tp
\gtrsim 4\,\ns$. Further manipulation of the stabilizing pulse
amplitude, indeed, enhances overall controllability of the
switching.
%------------------------------------------------------------------------------
\begin{center}
  \begin{figure}[t]
    \label{CoCuPy_sd5}
    \includegraphics[angle=0,width=0.99\columnwidth]{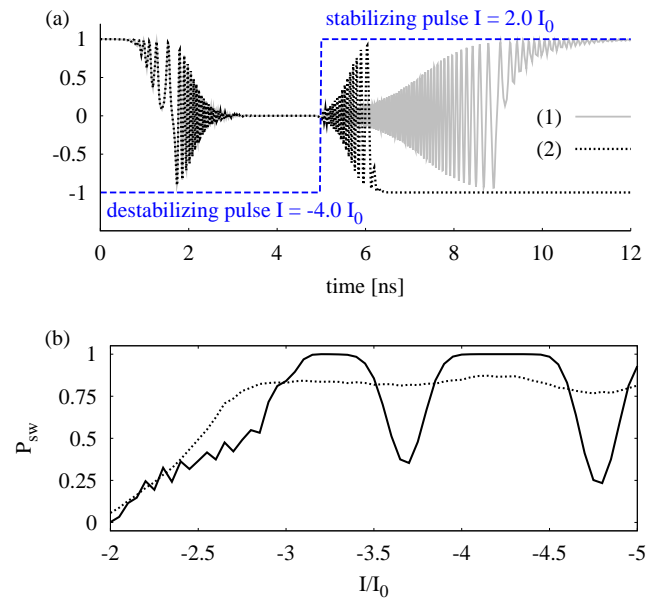}
    \caption{(Color online)
      Demonstration of {\em double-pulse switching scheme}
      making use of second (stabilizing) pulse.
      (a)~Evolution of $s_z$ spin component. Line (1) corresponds to
       a single $\tp = 5\,\ns$ ({\em destabilizing})
      current pulse of amplitude $I = -4.0\, I_0 $
      in zero magnetic field and zero temperature.
      Line (2) shows the evolution of $s_z$ in the case when the first pulse is
      followed by a second ({\em stabilizing}) pulse of opposite direction
      and $I = 2.0\, I_0$ (the double current pulse
      is shown by the blue dashed line).
      (b)~Thermally assisted switching probability $\Psw$  from the P to AP
      states, driven by a $5\,\ns$ ({\em destabilizing})
      current pulse  (negative) followed by a {\em stabilization} current pulse
       of $I = 2.0\, I_0$ and $t_{\rm p} \rightarrow \infty$,
      calculated as a function of the reduced current density $I/I_0$ of
      the {\em destabilizing} pulse, $I_0 = 10^8\,{\rm A cm^{-2}}$,
      for $T = 4.2\,\K$ (solid line) and $T = 77\,\K$ (dotted line).}
  \end{figure}
\end{center}
%------------------------------------------------------------------------------

In addition to the enhanced  controllability due to the
stabilizing pulse, we have observed enhancement of the switching
probability at  finite temperatures. In Fig.~5(b) we show the
switching probability as a function of the current density of
$5\,\ns$ destabilizing pulse, that is followed by a stabilization
current pulse of $I = 2.0\, I_0$ and $t_{\rm p} \rightarrow
\infty$. For $T = 4.2\,\K$ and  $I < -3\, I_0$, the probability
$\Psw$ oscillates with increasing current magnitude, similarly as
in the case of a single pulse, see Fig.~4(c). The regions of
successful switching are then broadened and the corresponding
amplitude is close to unity. In the case of $T = 77\,\K$, the
switching probability in this region is roughly constant and
approaches $\Psw \simeq 0.8$.

%----------------------------------------------
\section{Summary and conclusions}
%----------------------------------------------

We have studied dynamics of the current-pulse induced
magnetization switching in both standard and nonstandard spin
valves. The calculated switching diagrams reveal the pulse
parameters that are suitable for  ultra-fast spin switching. We
showed that the schemes of optimal switching  strongly depend on
the type of pillar structure, and are different for standard and
nonstandard spin valves.
The {\it wavy}-like torque in nonstandard spin valves introduces
stable points in the phase portrait for negative currents. These
stable points are responsible for the P/AP bistability in the
pulse switching diagram, which hinders obtaining successful
switching. Therefore, we proposed a switching scheme making use of
two current pulses, which can efficiently overcome the bistable
behavior. Additionally, the application of second, positive,
current pulse speeds-up the spin dynamics and fasten the switching
process for several times. We have also shown, that the proposed
scheme leads to enhanced switching probability even at high
temperatures.

The work has been supported by the the EU through the Marie Curie
Training network SPINSWITCH (MRTN-CT-2006-035327). 
M.~G. also acknowledges support within research projects 
MVTS POL/SR/UPJS07 and VEGA 1/0128/08. 
J.~B. also acknowledges support from the
Polish Ministry of Science and Higher Education as a research
project in years 2006 -- 2009 and National Scientific network
ARTMAG.
P.~B. also thanks to E.~Jaromirska and L.~L\'opez-D\'{\i}az for 
helpful discussions.

%\bibliography{cims-stt}

\end{document}